\documentclass[a4paper,11pt]{article}
\usepackage{jheppub}
\usepackage[T1]{fontenc} % if needed
\usepackage{amssymb}
\usepackage{amsmath}
\usepackage{xspace}
\usepackage{caption}
\usepackage{appendix}
\usepackage{float}
\usepackage{hyperref}
\usepackage{diagbox}
\usepackage{soul}
\usepackage{subfig}
\usepackage{slashed}
\usepackage{amsfonts}
\usepackage{multirow}
\usepackage{mathrsfs}
\usepackage{graphicx}
\usepackage{amsmath}
\usepackage{amssymb}
\usepackage{bm}
\usepackage{bbm}
\usepackage{color}
\usepackage{tcolorbox,colortbl}
\usepackage{booktabs}
\usepackage{comment}

\newcommand{\bea}{\begin{align}}
	\newcommand{\eea}{\end{align}}
\newcommand{\beq}{\begin{equation}}
	\newcommand{\eeq}{\end{equation}}
\newcommand{\bqa}{\begin{eqnarray}}
	\newcommand{\eqa}{\end{eqnarray}}

\newcommand{\M}{\ensuremath \text{M}}

\newcommand{\eps}{\epsilon}

\newcommand{\tekst}{\textrm}

\newcommand{\Li}{\tekst{Li}}

\newcommand{\als}{\alpha_s}

\allowdisplaybreaks[4]
\title{Next-to-leading order QCD corrections  to fully charm tetraquark hadronic decay}
\author[a,b]{Yefan Wang,}
\author[a,b]{Fengxiang Zhao,}
\author[a,b]{Ruilin Zhu}

\affiliation[a]{Department of Physics and Institute of Theoretical Physics, Nanjing Normal University,
Nanjing, Jiangsu 210023, China}
\affiliation[b]{Nanjing Key Laboratory of Particle Physics and Astrophysics, Nanjing Normal University, Nanjing,
Jiangsu 210023, China}

\abstract
{
We compute the next-to-leading order (NLO) QCD corrections to the light hadron decays of fully charm tetraquarks within the nonrelativistic QCD (NRQCD) factorization framework. The short-distance coefficients for the \(gg\) and \(q\bar{q}\) final states from fully charm tetraquarks are obtained analytically. The NLO corrections are found to be significant, altering the LO predictions by about \(170\%\) for the \(0^{++}\) state and \(30\%\) for the \(2^{++}\) state. The resulting \(\mathcal{R}_{\mathrm{LH}}\) values are of order \(10^{-5}\) to \(10^{-4}\) MeV, about three orders of magnitude larger than the diphoton channel, as expected for strong-interaction decays. Our results provide updated theoretical predictions for future experimental studies of fully charm tetraquark decays.
}

\emailAdd{wangyefan@nnu.edu.cn}
\emailAdd{ZhaoFengxiang@njnu.edu.cn}
\emailAdd{rlzhu@njnu.edu.cn}

\begin{document}
	\maketitle
	\flushbottom

\section{Introduction}
The past two decades have witnessed a revolution in hadron spectroscopy. Since the discovery of the X(3872) by the Belle Collaboration in 2003~\cite{Choi:2003ue,Acosta:2003zx}, a vast landscape of exotic hadrons—states that do not fit into the conventional quark-antiquark or three-quark pictures—has emerged. Among these, the fully charm tetraquark candidates, consisting of four charm quarks, occupy a particularly intriguing position. Unlike light or heavy-light exotics, fully heavy systems are expected to be relatively compact due to the absence of light quarks, which may render them more amenable to theoretical treatments based on heavy quark symmetry and nonrelativistic QCD. Consequently, they serve as important laboratories for testing various theoretical approaches to exotic hadrons, ranging from quark potential models to lattice QCD.

The experimental journey toward establishing fully charm tetraquarks began in 2020, when the LHCb Collaboration reported a narrow peak around 6900~MeV in the \(J/\psi\)-pair invariant mass distribution~\cite{LHCb:2020bwg}. This observation was subsequently confirmed by the ATLAS and CMS Collaborations~\cite{ATLAS:2023bft,CMS:2023owd}, lending strong support to the existence of a resonance in the fully charm sector. In the same year, the Belle Collaboration reported a possible peak at \(6267 \pm 43\)~MeV with a width of \(121 \pm 72\)~MeV~\cite{Belle:2023gln}. More recently, the CMS Collaboration, using a data sample corresponding to an integrated luminosity of \(135~\text{fb}^{-1}\) at \(\sqrt{s}=13\)~TeV, not only confirmed the 6900~MeV peak but also identified two additional structures at 6600~MeV and 7100~MeV, and determined their spin-parity quantum numbers~\cite{CMS:2025fpt,CMS:2026tiu}. Furthermore, the ATLAS Collaboration has observed the 6900~MeV peak in the \(J/\psi\)-\(\psi(2S)\) invariant mass distribution~\cite{ATLAS:2025nsd}, providing complementary evidence for this exotic state. These experimental advances have placed the study of fully charm tetraquarks at the forefront of modern hadron physics.

The theoretical exploration of four-charm bound states has a long history, dating back to the 1970s--1980s when Iwasaki and Chao first speculated on the possible existence of fully charm tetraquark near threshold~\cite{Iwasaki:1975pv,Chao:1980dv}. In the wake of the LHCb discovery, a proliferation of theoretical studies has emerged, employing a diverse array of methodologies. These include quark potential models~\cite{Anwar:2017toa,Karliner:2016zzc,Debastiani:2017msn,Wu:2016vtq,Liu:2019zuc,Jin:2020jfc,Liu:2021rtn,Faustov:2020qfm,Lu:2020cns,Wang:2019rdo}, the diquark-antidiquark picture~\cite{Bedolla:2019zwg,Zhu:2020xni,Giron:2020wpx}, gluonic tetraquark hybrid models~\cite{Tang:2024zvf,Tang:2024kmh,Tang:2025ept}, QCD sum rules~\cite{Chen:2016jxd,Zhang:2020xtb,Wang:2020dlo,Wang:2020ols,Albuquerque:2020hio,Wang:2021mma,Wan:2020fsk,Yang:2020wkh,Wang:2017jtz,Wang:2018poa,Agaev:2023wua,Agaev:2023gaq}, lattice QCD calculations~\cite{Li:2025vbd,Li:2025ftn,Meng:2024czd}, tetraquark-molecule mixing schemes~\cite{Santowsky:2021bhy}, Bethe-Salpeter equations~\cite{Li:2021ygk,Ke:2021iyh,Heupel:2012ua}, and dynamically generated resonance poles~\cite{Wang:2020wrp,Gong:2020bmg,Dong:2020nwy,Guo:2020pvt,Huang:2024jin}. Extensions to other fully heavy tetraquark systems, such as \(bb\bar{b}\bar{b}\) and \(cc\bar{b}\bar{b}\), have also been investigated~\cite{Esposito:2018cwh,Agaev:2025wyf,Wang:2025apq,Xia:2025mgk}. On the production side, the hadroproduction of fully charm tetraquarks has been studied extensively~\cite{Zhu:2020xni,Feng:2023agq,Belov:2024qyi,Wang:2025hex,Celiberto:2025vra,Celiberto:2025ziy,Zhang:2020hoh,Feng:2020riv}, including a complete next-to-leading order QCD calculation for their hadronic production~\cite{Wang:2025hex}. Fragmentation functions for quarks and gluons into fully charm tetraquarks have also been computed~\cite{Bai:2024flh,Bai:2024ezn,Celiberto:2024mab}. The very recent studies on fully charm tetraquark spectrum, decay and production properties can be seen in Refs.~\cite{Yang:2026rah,Wu:2026aqt,Chen:2026mrj,Wang:2026kcw,Celiberto:2026kks,Feng:2026orq,Yin:2026zbj,Anwar:2026ims,Paryev:2026lvf,Jia:2026onq}.  A concise review of the current status can be found in Ref.~\cite{Zhu:2024swp}.

Most existing studies on the decay properties of fully charm tetraquarks have focused on the \(J/\psi\)-pair channel~\cite{Sang:2023ncm,Chen:2024orv,Becchi:2020uvq,Zhang:2023ffe,Wang:2023kir,Biloshytskyi:2022dmo,Chen:2022sbf,Lu:2025lyu}, which is expected to be the dominant decay mode. More recently, the electromagnetic diphoton decay has also been investigated at NLO~\cite{Liu:2025mxv}. The light hadron decay modes—specifically, decays into two gluons (\(gg\)) and into light quark-antiquark pairs (\(q\bar{q}\) with \(q = u, d, s\))—have received comparatively less attention. A fully charm tetraquark, composed of \(c\bar{c}c\bar{c}\), must annihilate its constituent charm quarks in order to produce light partons. The associated Feynman diagrams involve multiple gluon exchanges and are therefore suppressed by the Okubo-Zweig-Iizuka (OZI) rule. Consequently, these decay channels are expected to have small branching fractions. Nonetheless, a quantitative understanding of these suppressed decay modes is of interest. They contribute to the total decay width of the tetraquark, and their relative magnitudes may provide information on the tetraquark's internal color structure. Furthermore, the calculation of these OZI-suppressed processes serves as a test of the NRQCD factorization framework when applied to multi heavy systems~\cite{Bodwin:1994jh}.

Previous calculations of the light hadron decays of fully charm tetraquarks have been carried out at leading order (LO) in the strong coupling constant~\cite{Sang:2023ncm}. In this work, we extend these calculations to next-to-leading order (NLO) by computing the short-distance coefficients for the decays \(T_{4c} \to gg\) and \(T_{4c} \to q\bar{q}\) within the NRQCD factorization framework. We consider both the \(J^{PC} = 0^{++}\) and \(J^{PC} = 2^{++}\) states, which are the most promising candidates from the experimental observations. The scale dependence of the short-distance coefficients and the impact of the NLO corrections on the numerical results are examined in detail.

The remainder of this paper is organized as follows. In Sec.~\ref{sec:formula}, we present the NRQCD factorization formulas for the decays \(T_{4c} \to gg\) and \(T_{4c} \to q\bar{q}\). The NLO calculation of the short-distance coefficients is described in Sec.~\ref{sec:analytical}. Numerical results for $\mathcal{R}_{\mathrm{LH}}(T_{4c})$ and their uncertainties are given and discussed in Sec.~\ref{sec:numerical}. Finally, we summarize our findings in Sec.~\ref{sec:conclusion}.

\section{Formalism}
\label{sec:formula}
In this work, we calculate the NLO QCD corrections to the total decay width of the $T_{4c}$ state decaying into light hadrons (LH), including both gluon and light-quark final-state contributions. 

The NRQCD factorization formula for $T_{4c}\rightarrow$ LH at the lowest
order in velocity can be expressed as \cite{Sang:2023ncm}  
\begin{align}
\Gamma(T^{0++}_{4c} \rightarrow \text{LH}) &= \frac{m_H}{128 m_c^4}\bigg( c_{1}\left\langle\mathcal{O}^{(0)}_{6\otimes\bar{6}}\right\rangle+c_{2}\left\langle\mathcal{O}^{(0)}_{\bar{3}\otimes3}\right\rangle \bigg.+c_{\text{mix}}\left\langle\mathcal{O}^{(0)}_{\text{mix}}\right\rangle
\bigg),
\nonumber\\
\Gamma(T^{2++}_{4c} \rightarrow \text{LH}) &= \frac{m_H}{128 m_c^4} c_{3}\left\langle\mathcal{O}^{(2)}_{\bar{3}\otimes3}\right\rangle,
\label{eq:fac}
\end{align}
where the superscript in $T_{4c}$ denotes the quantum number $J^{PC}$, $m_c$ and $m_H$ are the masses of the charm quark and $T_{4c}$, respectively. The coefficients $\{c_i\}$ are the short distance coefficients (SDCs). $\left\langle\mathcal{O}^{(J)}\right\rangle$ are the long distance matrix elements (LDMEs). %whose explicit form can be found in \cite{}. 
In the NRQCD factorization, the non-perturbative long-distance effects are absorbed into the LDMEs. Then the SDCs can be perturbatively calculated in powers of $\als$. \begin{align}
c_{i} = \sum_{j=0}^{\infty} \left(\frac{\als}{\pi}\right)^j c_{i,j}.
\end{align}
The required $T_{4c}$ amplitude can be obtained by projecting the $c(p_1)\bar{c}(p_2)c(p_3)\bar{c}(p_4)$ amplitude with free charm quarks. For the spin-singlet diquark contribution, the replacements are
\begin{align}
&\bar{u}(p_1)X_1v(p_3)\bar{u}(p_2)X_2v(p_4)\rightarrow -\text{Tr}\left[\Pi_{0} X_2\Pi_{0} X_1^C\right]\mathcal{C}_{6\otimes\bar{6}}^{ab;cd},\nonumber\\
&\bar{u}(p_1)X_1v(p_4)\bar{u}(p_2)X_2v(p_3)\rightarrow \text{Tr}\left[\Pi_{0} X_2\Pi^{C}_{0} X_1^C\right]\mathcal{C}_{6\otimes\bar{6}}^{ab;cd},
\label{eq:rep1}
\end{align}
where the subscript \(C\) denotes the charge conjugate, the color projection tensor is 
\begin{align}
\mathcal{C}_{\mathbf{6} \otimes \overline{\mathbf{6}}}^{ab;cd} &= \frac{1}{2\sqrt{6}}\left(\delta_{ac}\delta_{bd}+\delta_{ad}\delta_{bc}\right).
\end{align}
The spin-singlet projector is~\cite{Qiao:2012hp,Qiao:2012vt} 
\begin{align}
\Pi_{0}= \frac{\left(\frac{\slashed{P}}{4}+m_c\right)\gamma_{5}}{\sqrt{2}}.
\end{align}
where $P$ is the momentum of $T_{4c}$ and $P^2=m_H^2$. At the lowest order in the velocity expansion, the momenta are taken as
\begin{align}
p_1=p_2=p_3=p_4=\frac{P}{4}.
\end{align}

Then the spin-triplet diquark contribution can be obtained by the replacements 
\begin{align}
&\bar{u}(p_1)X_1v(p_3)\bar{u}(p_2)X_2v(p_4)\rightarrow -\text{Tr}\left[\Pi_{1\mu} X_2\Pi_{1\nu} X_1^C\right]\mathcal{C}_{\bar{3}\otimes3}^{ab;cd}J^{\mu \nu}_{0,2},\nonumber\\
&\bar{u}(p_1)X_1v(p_4)\bar{u}(p_2)X_2v(p_3)\rightarrow \text{Tr}\left[\Pi_{1\mu} X_2\Pi^{C}_{1\nu} X_1^C\right]\mathcal{C}_{\bar{3}\otimes3}^{ab;cd}J^{\mu \nu}_{0,2},
\label{eq:rep2}
\end{align}
where the color projection tensor is 
\begin{align}
\mathcal{C}_{\bar{3}\otimes3}^{ab;cd} &= \frac{1}{2\sqrt{3}}\left(\delta_{ac}\delta_{bd}-\delta_{ad}\delta_{bc}\right).   
\end{align}
and the spin-triplet projector is
\begin{align}
\Pi_{1,\mu}= \frac{\left(\frac{\slashed{P}}{4}+m_c\right)\gamma_{\mu}}{\sqrt{2}}.
\end{align}
The covariant projectors $J^{\mu \nu}_{0,2}$ in $D$ dimensions are
\begin{align}
J^{\mu \nu}_{0}&=\frac{1}{\sqrt{D-1}}\eta^{\mu \nu},\nonumber\\
J^{\mu \nu}_{2}&=\epsilon_{H,\alpha\beta}\left[\frac{1}{2}\eta^{\mu \alpha}\eta^{\nu \beta}+\frac{1}{2}\eta^{\mu \beta}\eta^{\nu \alpha}-\frac{1}{D-1}\eta^{\mu \nu}\eta^{\alpha\beta}\right],
\end{align}
where
\begin{align}
\eta^{\mu \nu} = -g^{\mu\nu} + \frac{P^\mu P^\nu}{m_H^2}.
\end{align}

\begin{figure}[ht]
	\centering
\begin{minipage}{0.3\linewidth}
	\centering
	\includegraphics[width=0.7\linewidth]{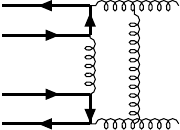}
 \caption*{(a)}
\end{minipage}
 \begin{minipage}{0.3\linewidth}
		\centering
    \includegraphics[width=0.7\linewidth]{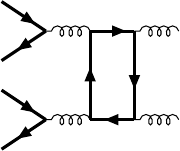}
   \caption*{(b)}
\end{minipage}
 \begin{minipage}{0.3\linewidth}
		\centering
    \includegraphics[width=0.7\linewidth]{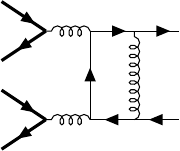}
   \caption*{(c)}
\end{minipage}
 \begin{minipage}{0.3\linewidth}
		\centering
    \includegraphics[width=0.7\linewidth]{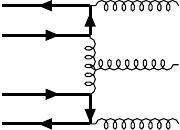}
   \caption*{(d)}
\end{minipage}
 \begin{minipage}{0.3\linewidth}
		\centering
    \includegraphics[width=0.6\linewidth]{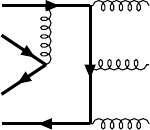}
   \caption*{(e)}
\end{minipage}
 \begin{minipage}{0.3\linewidth}
		\centering
    \includegraphics[width=0.6\linewidth]{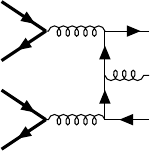}
   \caption*{(f)}
\end{minipage}
\caption{Sample Feynman diagrams contributing to $T_{4c}\rightarrow \text{LH}$ at NLO, the thick lines denote the massive charm quarks.}
\label{NLO}
\end{figure}

The one-loop virtual corrections to $T_{4c}\rightarrow{\text{LH}}$ have been calculated in our previous work of $T_{4c}$ production \cite{Wang:2025hex}. To obtain the complete NLO QCD corrections of $T_{4c}\rightarrow{\text{LH}}$, the real corrections with the processes $T_{4c}\rightarrow{ggg}$ and $T_{4c}\rightarrow{q\bar{q}}g$ need to be considered. Typical NLO QCD diagrams can be seen in figure \ref{NLO}, where diagrams (a)-(c) represent the one-loop virtual corrections, while (d)-(f) correspond to the real emission processes.

The IR divergences arising from the phase-space integrations of the amplitude squared $|\mathcal{M}(T_{4c}\rightarrow{ggg})|^2$ and $|\mathcal{M}(T_{4c}\rightarrow{gq\bar{q}})|^2$ need to be properly regularized. Thus the calculations need to be performed in $D$ dimensions. In this work we employ the reverse unitarity relation, in which the phase space integration can be converted to two-loop integrals via
\begin{align}
(2\pi)\delta(K_i^2)\theta(K^0_i) = \frac{i}{K_i^2+i\eps}
+\frac{-i}{K_i^2-i\eps}.
\end{align}
where $K_i$ is the momentum of the final state particle (gluon or light quark). In this way the $\delta$-functions are replaced by cut propagators, allowing the phase-space integrals to be treated within the framework of loop integration.
Then we can obtain several two-loop scalar integrals (SIs), which can be reduced to a set of basis integrals called master integrals (MIs) using the integration by parts (IBP) identities with the help of the package {\tt Kira} \cite{Klappert:2020nbg}. In this procedure the package {\tt CalcLoop}\footnote{\url{https://gitlab.com/multiloop-pku/calcloop}} has been used to shift the loop momenta and classify the SIs into several integral families. 

\section{Analytical calculations of SDCs}
\label{sec:analytical}
\begin{figure}[ht]
	\centering
	\begin{minipage}{0.2\linewidth}
		\centering
		\includegraphics[width=1\linewidth]{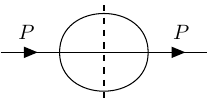}
		\caption*{$\M_{1}$}
	\end{minipage}
	\begin{minipage}{0.2\linewidth}
		\centering
		\includegraphics[width=1.15\linewidth]{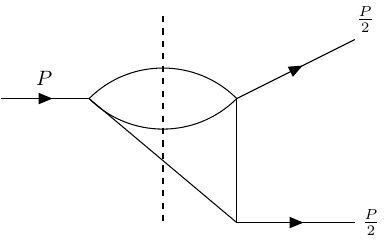}
		\caption*{$\M_{2}$}
	\end{minipage}
	\begin{minipage}{0.2\linewidth}
		\centering
		\includegraphics[width=1\linewidth]{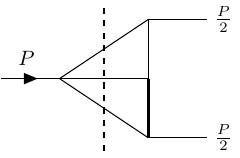}
		\caption*{$\M_{3}$}
	\end{minipage}
    	\begin{minipage}{0.2\linewidth}
		\centering
		\includegraphics[width=1\linewidth]{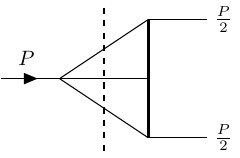}
		\caption*{$\M_{4}$}
	\end{minipage}
    	\begin{minipage}{0.2\linewidth}
		\centering
		\includegraphics[width=1.15\linewidth]{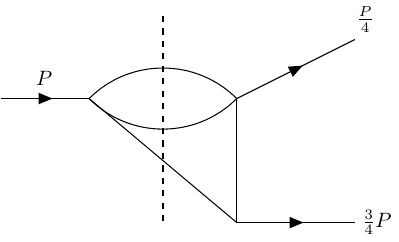}
		\caption*{$\M_{5}$}
	\end{minipage}
    	\begin{minipage}{0.2\linewidth}
		\centering
		\includegraphics[width=1\linewidth]{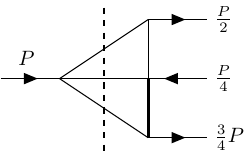}
		\caption*{$\M_{6}$}
	\end{minipage}
    	\begin{minipage}{0.2\linewidth}
		\centering
		\includegraphics[width=1\linewidth]{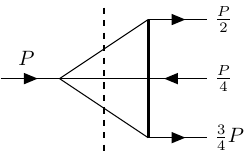}
		\caption*{$\M_{7}$}
	\end{minipage}
\caption{The topologies of the Master integrals. The thick black lines stand for the massive charm quark. The dashed propagators denote cut propagators.}
\label{Topo}
\end{figure}

After IBP reduction, we identify seven master integrals, each containing exactly three cut propagators, corresponding to the on-shell conditions of the final-state gluons or light quarks. These MIs can be expressed by
\begin{align}
\int{\mathcal D}^D q_1~{\mathcal D}^D q_2
\frac{1}{(D_{1})_\text{cut}~(D_{2})_\text{cut}~(D_{3})_\text{cut}~D_4^{n_4~}D_5^{n_5}~D_6^{n_6}~D_7^{n_7}~D_8^{n_8}},
\end{align}
where
\begin{align}
	{\mathcal D}^D q_i = \frac{\left(m_H^2 \right)^\epsilon}{i \pi^{D/2}\Gamma(1+\epsilon)}  d^D q_i.
\end{align}
The denominators read
\begin{align}
	D_1 &= q_1^2,&
	D_2 &= q_2^2,&
	D_3 &= (q_1+q_2+P)^2,\nonumber\\
	D_4 &= (q_1+\frac{P}{2})^2,&
	D_5 &= (q_2+\frac{P}{2})^2,&
	D_6 &= (q_1+\frac{P}{2})^2-\frac{m_H^2}{16},\nonumber\\
	D_7 &= (q_2+\frac{P}{2})^2-\frac{m_H^2}{16},&
	D_8 &= (q_1+\frac{3P}{4})^2-\frac{m_H^2}{16}.
\end{align}
And the subscript “cut” indicates that the relevant propagator is on shell
\begin{align}
\frac{1}{\left(Q_i^2+i\eps\right)_{\text{cut}}} = -2i\pi\delta(Q_i^2)\theta(Q^0_i),\end{align}
where $Q_i$ is the linear combination of loop momenta $q_1$, $q_2$ and external momentum $P$. Note that these eight denominators are not linearly independent; at most five denominators can simultaneously appear in the given integral. Since all MIs depend only on the single scale $m_H^2$, we cannot find a dimensionless variable to construct the differential equations. Instead we can introduce another scale as an auxiliary scale. For example, we replace the mass term in $D_6$ as
\begin{align}
 D_6 \rightarrow (q_1+\frac{P}{2})^2-m_x^2. 
\end{align}
Then we define $z \equiv m_H^2/m_x^2$. The physical result is recovered in the limit $z\rightarrow16$ after all calculations. This auxiliary parameter allows us to derive canonical differential equations \cite{Kotikov:1991pm,Henn:2013pwa} with respect to $z$. Then the differential equations
can be solved recursively in terms of multiple polylogarithms (MPLs) \cite{Goncharov:1998kja}, which are defined by $G(x)\equiv 1$ and
\bqa
	G(l_1,l_2,\ldots,l_n,x) &\equiv & \int_0^x \frac{\text{d} t}{t - l_1} G(l_2,\ldots,l_n,t)\, ,\\
	G(\overrightarrow{0}_n , x) & \equiv & \frac{1}{n!}\ln^n x\, .
\eqa
The number of elements in the set $\{l_1,l_2,\ldots,l_n\}$ is referred to as the transcendental $weight$ of the MPLs.
The boundary conditions can be fixed by the PSLQ algorithm \cite{Ferguson:1999aa} with the high-precision numerical results from the package {\tt AMFlow} \cite{Liu:2017jxz,Liu:2020kpc,Liu:2022chg,Liu:2022mfb}. Since the required MPLs are at most of weight two, they can be expressed entirely in terms of $\Li_2$ and 
$\log$ functions. The analytic expressions read:
\begin{align}
I_1=&\int\prod_{i=1}^2{\mathcal D}^D q_i   
\frac{1}{(D_{1})_\text{cut}(D_{2})_\text{cut}(D_{3})_\text{cut}} = m_H^2\left(-\frac{\pi }{2}-\frac{13 \pi}{4}\eps+\left(\frac{\pi ^3}{2}-\frac{115 \pi }{8} \right)\eps^2+ \mathcal{O}(\eps^3)\right),
\nonumber\\
I_2=&\int\prod_{i=1}^2{\mathcal D}^D q_i   
\frac{1}{(D_{1})_\text{cut}(D_{2})_\text{cut}(D_{3})_\text{cut}D_5} = 2\pi+\left(14\pi-\frac{\pi^3}{4}\right)\eps+ \mathcal{O}(\eps^2),
\nonumber\\
I_3=&\int\prod_{i=1}^2{\mathcal D}^D q_i   
\frac{1}{(D_{1})_\text{cut}(D_{2})_\text{cut}(D_{3})_\text{cut}D_4D_7} = \frac{1}{m_H^2}\left(-2\pi^3+ \mathcal{O}(\eps)\right),
\nonumber\\
I_4=&\int\prod_{i=1}^2{\mathcal D}^D q_i   
\frac{1}{(D_{1})_\text{cut}(D_{2})_\text{cut}(D_{3})_\text{cut}D_6D_7} = \frac{1}{m_H^2}\left(-\frac{8\pi^3}{3}+ \mathcal{O}(\eps)\right),
\nonumber\\
I_5=&\int\prod_{i=1}^2{\mathcal D}^D q_i   
\frac{1}{(D_{1})_\text{cut}(D_{2})_\text{cut}(D_{3})_\text{cut}D_8}= \frac{4\pi}{3}-\frac{8\pi  \log (2)}{9}+ \mathcal{O}(\eps),
\nonumber\\
I_6=&\int\prod_{i=1}^2{\mathcal D}^D q_i   
\frac{1}{(D_{1})_\text{cut}(D_{2})_\text{cut}(D_{3})_\text{cut}D_5D_8}= \frac{1}{m_H^2}\bigg(\frac{8\pi}{3}\left(\text{Li}_2\left(-\frac{1}{3}\right)-\text{Li}_2\left(\frac{1}{3}\right)\right)
\nonumber\\
&\quad\quad\quad+ 4\pi^3+\mathcal{O}(\eps)\bigg),
\nonumber\\
I_7=&\int\prod_{i=1}^2{\mathcal D}^D q_i   
\frac{1}{(D_{1})_\text{cut}(D_{2})_\text{cut}(D_{3})_\text{cut}D_7D_8}= \frac{1}{m_H^2}\bigg(-\frac{8\pi}{3}\text{Li}_2\left(\frac{1}{4}\right)-\frac{4 \pi ^3}{3}-\frac{16 \pi\log(2)^2  }{3}
\nonumber\\
&\quad\quad\quad+\frac{16 \pi\log(2)\log(3)  }{3}+\mathcal{O}(\eps)\bigg),
\end{align}
where the MIs are expanded to the required order in $\eps$ such that the overall accuracy is sufficient for the NLO calculation. And the topology diagrams of MIs are shown in Figure \ref{Topo}.

With all two-loop MIs at hand, we can obtain the analytical expressions of real corrections. Combining with the virtual corrections in our previous work, we find the IR divergences are completely canceled, which provides a solid validation of the correctness of our computation. Finally, the SDCs up to NLO can be expressed as 
\begin{align}
c_{1} &= \frac{484 \pi ^3 \als^4}{27 m_c^4 m_H}\left(1+\left(\frac{\als}{\pi}\right)X_{1,1}+\mathcal{O}(\als^2)\right),
\nonumber\\
c_{2} &= \frac{72 \pi ^3 \als^4}{m_c^4 m_H}\left(1+\left(\frac{\als}{\pi}\right)X_{2,1}+\mathcal{O}(\als^2)\right),
\nonumber\\
c_{mix} &= \frac{88\sqrt{6} \pi ^3 \als^4}{3m_c^4 m_H}\left(1+\left(\frac{\als}{\pi}\right)X_{mix,1}+\mathcal{O}(\als^2)\right),
\nonumber\\
c_{3} &= \frac{6272 \pi ^3 \als^4}{135 m_c^4 m_H}\left(1+\frac{48}{49}n_l+\left(\frac{\als}{\pi}\right)X_{3,1}+\mathcal{O}(\als^2)\right).
\end{align}
where $n_l$ is the number of light quark flavors. Note that at LO, the decay of $T^{0++}_{4c}\rightarrow q \bar{q}$ is forbidden due to helicity suppression. The analytical expressions of the $\{X_{i,1}\}$ read
\begin{align}
X_{1,1} &=\frac{31}{3}\log\left(\frac{\mu^2}{m_H^2}\right)
+\frac{1556479}{5808} \text{Li}_2\left(\frac{1}{3}\right)+\frac{35297}{2904}\text{Li}_2\left(-\frac{1}{3}\right)
+\frac{3689}{528}\bigg(
-\text{Li}_2\left(\frac{3}{5}\right)
\nonumber\\&\quad
-2\text{Li}_2\left(\frac{1}{5}\right)
+\text{Li}_2\left(-\frac{1}{5}\right)+\text{Li}_2\left(-\frac{3}{5}\right)-\text{Li}_2\left(-\frac{5}{3}\right)-\frac{1}{2}\log ^2(5)\bigg)
-\frac{2720629\pi^2}{69696}
\nonumber\\&\quad
+\frac{437870}{1089}
+\frac{5}{44} \log ^2\left(7-4 \sqrt{3}\right)
-\frac{989}{132} \log ^2\left(3+2 \sqrt{2}\right)+\frac{793247}{5808}\log ^2(3)
\nonumber\\&\quad
+\frac{1985\sqrt{2}}{99}\log \left(3+2 \sqrt{2}\right)
-\frac{175391}{594}\log (2)
+n_l\bigg[
-\frac{2}{3}\log \left(\frac{\mu^2}{m_H^2}\right)
\nonumber\\&\quad
+\frac{125}{99}\left(\text{Li}_2\left(\frac{1}{3}\right)-\text{Li}_2\left(-\frac{1}{3}\right)\right)-\frac{619 \pi ^2}{726}-\frac{76115}{3267}
+\frac{57766}{9801}{\log (2)}\bigg]
\nonumber\\
&\quad\approx
(10.3333-0.666667n_l)\log\left(\frac{\mu^2}{m_H^2}\right)+85.4819-26.7752n_l
\\
X_{2,1} &=\frac{31}{3}\log\left(\frac{\mu^2}{m_H^2}\right)+\frac{29881}{104976}\text{Li}_2\left(\frac{1}{3}\right)+\frac{361417}{437400}\text{Li}_2\left(-\frac{1}{3}\right)+\frac{7}{432}\bigg(
-\text{Li}_2\left(\frac{3}{5}\right)
\nonumber\\&\quad
-2\text{Li}_2\left(\frac{1}{5}\right)+\text{Li}_2\left(-\frac{1}{5}\right)-2\text{Li}_2\left(-\frac{5}{3}\right)-\log^2(5)+\log(3)\log(5)\bigg)
\nonumber\\&\quad
-\frac{64954453 \pi ^2}{31492800}+\frac{267974}{6561}-\frac{11}{108} \log ^2\left(7-4 \sqrt{3}\right)-\frac{41}{36} \log ^2\left(3+2\sqrt{2}\right)
\nonumber\\&\quad
+\frac{2830477}{5248800} \log ^2(3)+\frac{113}{243} \sqrt{2} \log \left(3+2\sqrt{2}\right)+\frac{4730339}{196830}\log (2)
\nonumber\\&\quad
+n_l\bigg[
-\frac{2}{3}\log \left(\frac{\mu^2}{m_H^2}\right)+\frac{55}{6561}\left(\text{Li}_2\left(\frac{1}{3}\right)-\text{Li}_2\left(-\frac{1}{3}\right)\right)+\frac{1073 \pi ^2}{4374}
\nonumber\\&\quad
-\frac{82067}{19683}-\frac{37946}{59049}\log (2)\bigg]
\nonumber\\
&\quad\approx 
\left(10.3333-0.666667n_l\right)\log\left(\frac{\mu^2}{m_H^2}\right)+34.5643-2.18806{n_l}
\\
X_{mix,1} &=\frac{31}{3}\log\left(\frac{\mu^2}{m_H^2}\right)+\frac{16639}{4752}\bigg(
-\text{Li}_2\left(\frac{3}{5}\right)-2\text{Li}_2\left(\frac{1}{5}\right)+\text{Li}_2\left(-\frac{1}{5}\right)-2\text{Li}_2\left(-\frac{5}{3}\right)
\nonumber\\&\quad
-\log^2(5)+\log(3)\log(5)\bigg)+\frac{53905}{3888}\text{Li}_2\left(\frac{1}{3}\right)+\frac{1223329}{178200}\text{Li}_2\left(-\frac{1}{3}\right)
\nonumber\\&\quad
-\frac{67040911 \pi ^2}{12830400}+\frac{228242}{2673}+\frac{7}{1188}\log ^2\left(7-4 \sqrt{3}\right)-\frac{1709}{396}\log ^2\left(3+2\sqrt{2}\right)
\nonumber\\&\quad
+\frac{1334209}{194400}\log ^2(3)+\frac{27419}{2673} \sqrt{2} \log \left(3+2\sqrt{2}\right)-\frac{398467}{80190}\log (2)
\nonumber\\&\quad
+n_l\bigg[
-\frac{2}{3}\log \left(\frac{\mu^2}{m_H^2}\right)+\frac{65}{243}\left(\text{Li}_2\left(\frac{1}{3}\right)-\text{Li}_2\left(-\frac{1}{3}\right)\right)+\frac{227 \pi ^2}{1782}
\nonumber\\&\quad
-\frac{69491}{8019}+\frac{2998}{24057}\log (2)\bigg]
\nonumber\\
&\quad\approx 
\left(10.3333-0.666667n_l\right)\log\left(\frac{\mu^2}{m_H^2}\right)+54.978-7.14155n_l
\\
X_{3,1} &= \frac{31}{3}\log\left(\frac{\mu^2}{m_H^2}\right)+\frac{7228441}{169344}\text{Li}_2\left(\frac{1}{3}\right)-\frac{155}{504}\text{Li}_2\left(-\frac{1}{3}\right)+\frac{3}{8}\log ^2\left(7-4 \sqrt{3}\right)
\nonumber\\&\quad
-\frac{545}{56}\log ^2\left(2 \sqrt{2}+3\right)+\frac{7176361}{338688}\log ^2\left(3\right)-\frac{75033925 \pi ^2}{2032128}-\frac{7}{\sqrt{3}} \log\left(4 \sqrt{3}+7\right)
\nonumber\\&\quad
+\frac{2021}{42 \sqrt{2}} \log\left(2 \sqrt{2}+3\right)-\frac{405929}{18144}\log(2)+\frac{13951639}{42336}
+n_l\bigg[
\frac{1390}{147}\log\left(\frac{\mu^2}{m_H^2}\right)
\nonumber\\&\quad
-\frac{56545}{21168}\text{Li}_2\left(\frac{1}{3}\right)-\frac{17873}{3024}\text{Li}_2\left(-\frac{1}{3}\right)-\frac{211}{196} \log ^2(3)
+\frac{45}{7} \bigg(\text{Li}_2\left(-\frac{5}{3}\right)+\text{Li}_2\left(\frac{1}{5}\right)
\nonumber\\&\quad
+\frac{1}{2}\text{Li}_2\left(\frac{3}{5}\right)-\frac{1}{2}\text{Li}_2\left(-\frac{1}{5}\right)+\frac{1}{2}\log ^2(5)-\frac{1}{2} \log (3) \log (5)\bigg)
+\frac{13781 \pi ^2}{28224}
\nonumber\\&\quad
+\frac{1361303}{47628}\log (2)-\frac{426121}{15876}
\bigg]+n_l^2\bigg[-\frac{32}{49}\log\left(\frac{\mu^2}{m_H^2}\right)-\frac{64}{49} \log(2)-\frac{160}{147}\bigg]
\nonumber\\
&\quad\approx \left(10.3333+9.45578 n_l-0.653061 n_l^2 \right)\log\left(\frac{\mu^2}{m_H^2}\right)+12.6093-3.72839 n_l
\nonumber\\
&\quad\quad-1.99377n_l^2
\end{align}

\section{Numerical results}
\label{sec:numerical}
\begin{figure}[ht]
	\centering
\includegraphics[width=1\linewidth]{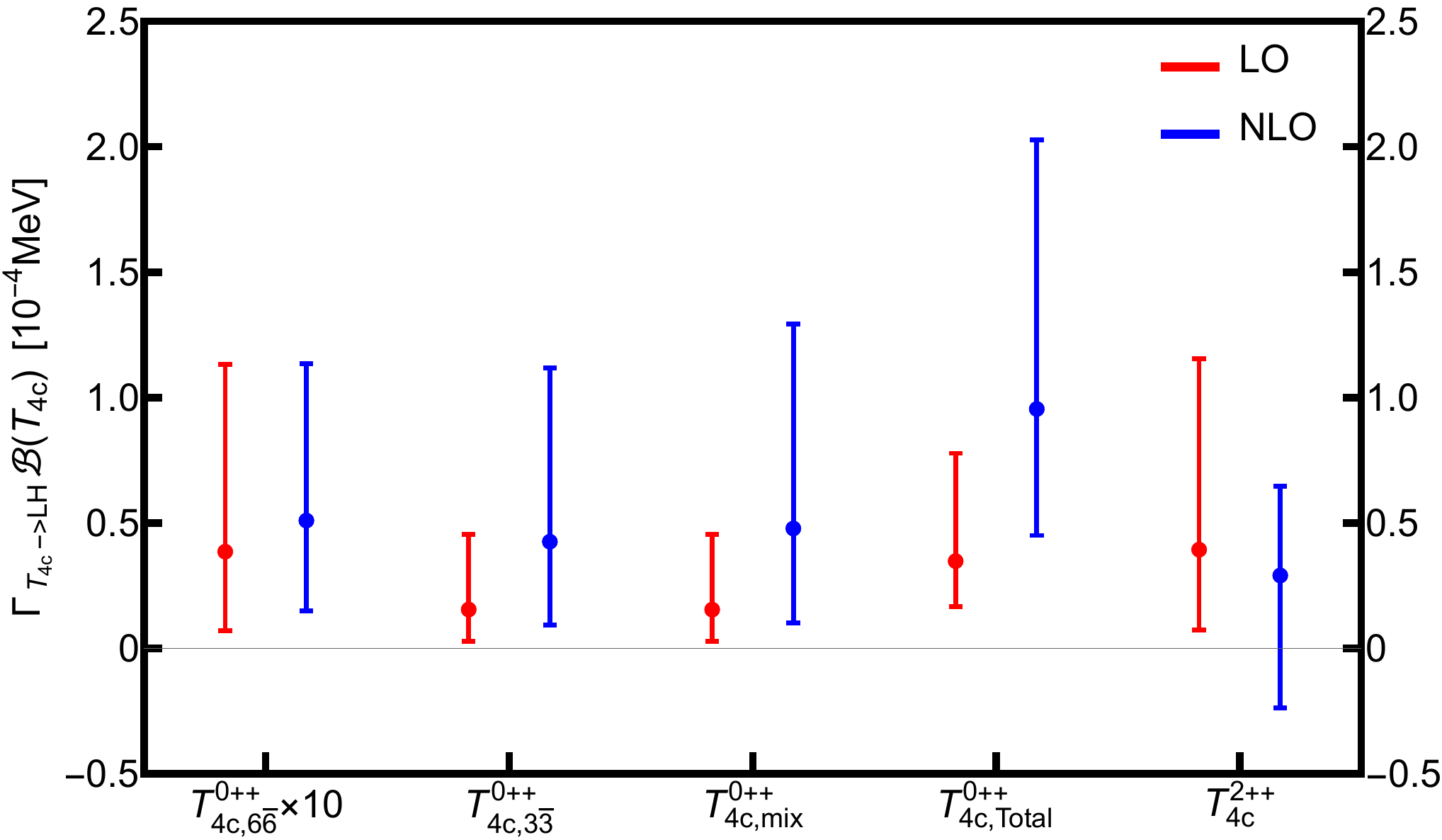}
\caption{The numerical values of $\mathcal{R}_{\text{LH}}(T^{0++}_{4c})$ and $\mathcal{R}_{\text{LH}}(T^{2++}_{4c})$. The error bars denote the uncertainties.}
\label{plot}
\end{figure}

Using the analytical expressions derived above, we present the numerical results. We take the input parameters as
\begin{align}
&m_H = 4m_c = 6847^{+44+48}_{-28-20} \,\text{MeV}\quad
\nonumber\\
&m_Z =   91.1876 \,\text{GeV},
\quad
\als(m_Z) = 0.1181, \quad n_l=3,
\end{align}
where we take the mass of $X(6900)$ as the $m_H$ from the latest CMS measurement \cite{CMS:2023owd}. Then the package {\tt{RunDec}} \cite{Herren:2017osy} was used to obtain $\als$ at other scales. The strong coupling $\als$ at $2m_c$, $4m_c$ and $8m_c$ is evaluated to be 
\begin{align}
&\als(2m_c) = 0.242997,\quad    \als(4m_c) = 0.195959,\quad \als(8m_c) = 0.166230.
\end{align}
To calculate the decay width of $T_{4c}\rightarrow \text{LH}$, we also need the LDMEs as in Eq. (\ref{eq:fac}). In this paper we adopt the LDMEs from our previous works \cite{Wang:2025hex,Liu:2025mxv},
\begin{align}
\langle\mathcal{O}_{\bar{3}\otimes3}^{(2)}\rangle\mathcal{B}(T_{4c}) =4.44^{+1.60+2.58+1.24}_{-1.60-1.14-0.00}\times10^{-5}\text{GeV}^9,
\end{align}
where $\mathcal{B}(T_{4c})$ stands for the branch ratio of $T_{4c}\rightarrow2J/\psi$.
The uncertainties are from the experimental uncertainty, scale uncertainty and parton distribution function uncertainty. More detail can be found in \cite{Wang:2025hex}. In the absence of direct experimental measurements for the 0++ fully charm tetraquark, we make the following assumption,
\begin{align}
\langle\mathcal{O}_{6\otimes\bar{6}}^{(0)}\rangle =  \langle\mathcal{O}_{\bar{3}\otimes3}^{(0)}\rangle =  \langle\mathcal{O}_{mix}^{(0)}\rangle
= \langle\mathcal{O}_{\bar{3}\otimes3}^{(2)}\rangle = 4.44^{+1.60+2.58+1.24}_{-1.60-1.14-0.00}\times\frac{10^{-5}}{\mathcal{B}(T_{4c})}\text{GeV}^9.
\end{align}
The above equality corresponds to the mixing angle $\theta = \pi/4$ \cite{Zhang:2023ffe}. In the absence of experimental measurements for the branching fraction of $T_{4c}\rightarrow 2J/\psi$, for simplicity we define 
\begin{align}
\mathcal{R}_{\text{LH}}(T_{4c})\equiv\Gamma(T_{4c}\rightarrow\text{LH})\mathcal{B}(T_{4c}\rightarrow2J/\psi).
\end{align}
In Table \ref{tab:num}, we present the numerical results of $\mathcal{R}_{\text{LH}}(T_{4c})$ up to NLO at different renormalization scale $\mu$. The other input parameters, including the LDMEs and $m_H$, are chosen as their central values. Here we have $\mathcal{R}_{\text{LH}}(T^{0++}_{4c}) = \mathcal{R}_{\text{LH}}(T^{0++}_{6\otimes\bar{6},4c})+\mathcal{R}_{\text{LH}}(T^{0++}_{\bar{3}\otimes3,4c})+\mathcal{R}_{\text{LH}}(T^{0++}_{mix,4c})$. It can be seen that contributions from $T^{0++}_{6\otimes\bar{6},4c}$ are much smaller than others. The same feature is also observed in the $T^{0++}_{4c}\rightarrow \gamma\gamma$. Interestingly, the $T^{0++}_{mix,4c}$ contributions are similar to $T^{0++}_{\bar{3}\otimes3,4c}$ in both LO and NLO. The total $\mathcal{R}_{\text{LH}}$ for $T^{0++}_{4c}$ is similar to that for $T^{2++}_{4c}$ at LO, but the difference becomes more significant at NLO. At the typical scale $\mu=4m_c=m_H$, the NLO QCD correction increases the LO value by approximately $170\%$ for $T^{0++}_{4c}$. Conversely, for $T^{2++}_{4c}$ the NLO correction decreases the LO value by about $30\%$. Overall, the NLO QCD corrections are significant. And the scale dependence of the results is also significant. For instance, at $\mu=2m_c$, $\mathcal{R}_{\text{LH}}(T^{2++}_{4c})$ even becomes negative, suggesting that higher-order corrections may still be important. After including all uncertainties, we also plot the numerical results in Figure \ref{plot}. One can see that the uncertainties of 
$\mathcal{R}_{\text{LH}}$ increase significantly. The final $\mathcal{R}_{\text{LH}}$ values are found to be of order $10^{-5}$ to $10^{-4}$ MeV with $\mathcal{R}_{\text{LH}}(T^{0++}_{4c})$ exceeding $\mathcal{R}_{\text{LH}}(T^{2++}_{4c})$. These values are approximately three orders of magnitude higher than the corresponding results in $\mathcal{R}_{\gamma\gamma}$ \cite{Liu:2025mxv}. This is expected, as the light hadron decays proceed via the strong interaction.
\begin{table}[ht]
    \centering
        \begin{tabular}{|c|c|c|c|c|}
            \hline
            & [$10^{-5}$ MeV] & $\mu = 2m_c$ & $\mu = 4m_c$ & $\mu = 8m_c$ \\
            \hline
            \multirow{2}{*}{$\mathcal{R}_{\text{LH}}(T^{0++}_{6\otimes\bar{6},4c})$}
            & LO  &0.91 &0.39 &0.20  \\
            \cline{2-5}
            & NLO &0.46 &0.51 &0.38  \\
            \cline{2-5}
            \hline
            \multirow{2}{*}{$\mathcal{R}_{\text{LH}}(T^{0++}_{\bar{3}\otimes3,4c})$} 
            & LO  &3.66  &1.55 &0.80 \\
            \cline{2-5}
            & NLO &8.32  &4.25 &2.48 \\
            \cline{2-5}
            \hline
            \multirow{2}{*}{$\mathcal{R}_{\text{LH}}(T^{0++}_{mix,4c})$} 
            & LO  &3.66  &1.55 &0.80  \\
            \cline{2-5}
            & NLO &9.88  &4.78 &2.71  \\
            \cline{2-5}
            \hline
            \multirow{2}{*}{$\mathcal{R}_{\text{LH}}(T^{0++}_{4c})$}
            & LO  &8.23  &3.48  &1.80 \\
            \cline{2-5}
            \cline{2-5}
            & NLO &18.66  &9.55  &5.57  \\
            \cline{2-5}
            \hline
            \multirow{2}{*}{$\mathcal{R}_{\text{LH}}(T^{2++}_{4c})$}
            & LO  &9.31&  3.94&  2.04\\
            \cline{2-5}
            & NLO &-2.03& 2.91& 2.83\\
            \cline{2-5}
            \hline
        \end{tabular}
\caption{The numerical values of  $\mathcal{R}_{\text{LH}}(T^{0++}_{4c})$ and $\mathcal{R}_{\text{LH}}(T^{2++}_{4c})$ at different scales. The other input parameters are set to their central values.} 
\label{tab:num}
\end{table}
\section{Conclusion}
\label{sec:conclusion}
In this work, we have calculated the NLO QCD corrections to the light hadron decays of fully charm tetraquarks within the NRQCD factorization framework. The short-distance coefficients for \(T_{4c} \to gg\) and \(T_{4c} \to q\bar{q}\) are obtained analytically, and the scale dependence is examined in detail. The NLO corrections are found to be significant: the \(0^{++}\) result is enhanced by about \(170\%\) at \(\mu = m_H\), while the \(2^{++}\) result is reduced by about \(30\%\), with a negative value appearing at \(\mu = m_H/2\). The \(\bar{3}\otimes 3\) and mixing diquark configurations dominate the \(0^{++}\) channel, similar to the diphoton case. The final \(\mathcal{R}_{\mathrm{LH}}\) values are of order $10^{-5}$ to $10^{-4}$ MeV, about three orders of magnitude larger than the diphoton channel, as expected for strong-interaction decays. These results provide useful theoretical inputs for future experimental studies of fully charm tetraquark decays.

\section*{Acknowledgments}
We thank Zhe Li for helpful discussions about Feynman integrals. This work is supported by
the National Natural Science Foundation of China
Grants No.12322503, No.12405117 and No.12675120.
\bibliographystyle{JHEP}
\bibliography{ref}
\end{document}